\documentclass[pre,showpacs,preprint,tightenlines,nofootinbib]{revtex4}

\usepackage{graphicx,bm,amsfonts}
\usepackage[tbtags]{amsmath}

\newcommand{\ij}{i\kern -0.08em j}

\def\beq{\begin{equation}}
\def\eeq{\end{equation}}
\def\eeql#1{\label{#1} \end{equation}}

\newcommand{\To}{\rightarrow}

\newcommand{\e}{\mathrm{e}}
\newcommand{\pt}{\partial}

\newcommand{\Oc}{\mathcal{O}}

\newcommand{\Ic}{\mathcal{I}}

\begin{document}

\title{Galvanic coupling of flux qubits:\\ simple theory and tunability}
\author{Alec \surname{Maassen van den Brink}}
\email{alec@dwavesys.com}
\affiliation{D-Wave Systems Inc., 100-4401 Still Creek Drive, Burnaby, B.C., V5C 6G9 Canada}

\date{\today}

\begin{abstract}
Galvanic coupling of small-area (three-junction) flux qubits, using shared large Josephson junctions, has been shown to yield appreciable interaction strengths in a flexible design, which does not compromise the junctions' intrinsic good coherence properties. For an introduction, I recapitulate an elementary derivation of the coupling strength, which is subsequently generalized to the case of tunable coupling for a current-biased shared junction. While the ability to vary coupling constants by, say, 20\% would be useful in experiments, \emph{sign}-tunability (implying switchability) is highly preferable for several quantum-computing paradigms. This note sketches two ideas: a ``crossbar'' design with competing ferro- and antiferromagnetic current-biased tunable couplings, and a ``mediated'' one involving an extra loop between the qubits. The latter is a variation on proposals for tunable capacitive coupling of charge qubits, and tunable inductive coupling of large-area flux qubits.
\end{abstract}

\maketitle

The design philosophy of the single three-junction (3JJ) qubit is to obtain a bistable system without relying on magnetic energy, by using multiple Josephson junctions~\cite{orlando}. This revives as old idea for using multistable multi-junction loops~\cite{yama} and generalizes it to the quantum regime, in the presence of an arbitrary (but typically nearly half-integer) flux bias. As shown by Levitov \emph{et al.}~\cite{levitov} and apparently rediscovered by Butcher~\cite{butcher}, the coupling of several such qubits can be implemented analogously, using large Josephson junctions inserted into the shared legs of adjacent qubits. In this way, an appreciable coupling strength should be easier to achieve than using inductive coupling, especially since the 3JJ loops typically have a small area (if anything, for a circuit of this type the design challenge will be to avoid the coupling strength being too large). In Levitov \emph{et al.}'s preprint, only the order of magnitude of the coupling strength is estimated (correctly). In Butcher's report (Section~3.5.1), it is stated that the Hamiltonians for Josephson and inductive coupling are identical. However, the former is not given let alone derived, while the correct form for the latter~\cite{H_2qb} has eluded the Delft group (overestimation by a factor two) both in Butcher's thesis [Eq.~(3.3)] and for some time thereafter~\cite{paauw}.

In this note, therefore, first the coupling strength will be derived, recapitulating a result meanwhile published in~\cite{galv-expt}. Both the result and the derivation turn out to be very similar to the inductive case, confirming Butcher's statement. However, the calculation is decidedly easier, since the leading answer is found by studying the classical potential for vanishing inductance (as opposed to expanding the full Hamiltonian to first order in the inductances).

\setlength{\unitlength}{1mm}
\begin{figure}[ht]
\begin{picture}(70,50)
  \put(7,5){\line(1,0){56}}
  \put(7,45){\line(1,0){56}}
  \put(7,5){\line(0,1){40}}
  \put(63,5){\line(0,1){40}}
  \put(35,5){\line(0,1){40}}
  \put(31,21){\line(1,1){8}}
  \put(31,29){\line(1,-1){8}}
  \put(18,2){\line(1,1){6}}
  \put(18,8){\line(1,-1){6}}
  \put(18,42){\line(1,1){6}}
  \put(18,48){\line(1,-1){6}}
  \put(46,2){\line(1,1){6}}
  \put(46,8){\line(1,-1){6}}
  \put(46,42){\line(1,1){6}}
  \put(46,48){\line(1,-1){6}}
  \put(5,23){\line(1,1){4}}
  \put(5,27){\line(1,-1){4}}
  \put(61,23){\line(1,1){4}}
  \put(61,27){\line(1,-1){4}}
  \put(20,23.8){$\phi_\mathrm{x}^a$}
  \put(48,23.8){$\phi_\mathrm{x}^b$}
  \put(30,23.8){0}
  \put(20,47.5){1}
  \put(3,23.8){2}
  \put(20,0){3}
  \put(48,47.5){4}
  \put(65,23.8){5}
  \put(48,0){6}
\end{picture}
\caption{Two Josephson-coupled 3JJ qubits.}
\label{circuit}
\end{figure}
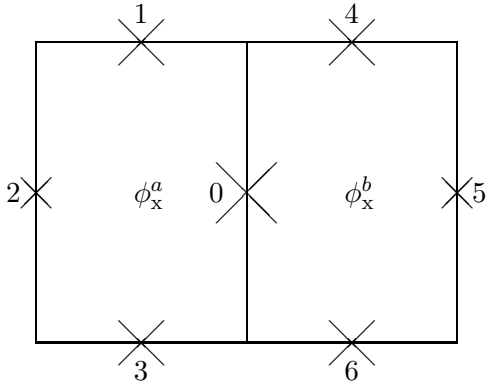

The circuit is shown in Fig.~\ref{circuit}. The relevant part of the Hamiltonian reads simply\linebreak $U_\mathrm{J}=\sum_{j=0}^6-E_j\cos\phi_j$. For vanishing inductances, this is subject to flux quantization $\phi_1+\phi_2+\phi_3+\phi_0=\phi_\mathrm{x}^a$ and $\phi_4+\phi_5+\phi_6-\phi_0=\phi_\mathrm{x}^b$ (note the minus sign in the latter relation), where $\phi_\mathrm{x}^{a,b}=2\pi\Phi_\mathrm{x}^{a,b}/\Phi_0$ are external flux biases in phase units. We focus on the simplest case $\phi_\mathrm{x}^{a,b}=\pi$, $E_{1,3,4,6}=E$, $E_{2,5}=\alpha E$ ($\frac{1}{2}<\alpha<1$), $E_0\gg E$, so that the potential becomes
\beq\begin{split}
  U_\mathrm{J}(\bm{\phi})&=-E_0\cos\phi_0
  +E[-\cos\phi_1-\cos\phi_3+\alpha\cos(\phi_1{+}\phi_3{+}\phi_0)\\
  &\quad\,-\cos\phi_4-\cos\phi_6+\alpha\cos(\phi_4{+}\phi_6{-}\phi_0)]\;.
\end{split}\eeql{UJ}
The wells, corresponding to the classical qubit states, are partly characterized by $\pt_0U_\mathrm{J}=\pt_1U_\mathrm{J}=\pt_3U_\mathrm{J}=\pt_4U_\mathrm{J}=\pt_6U_\mathrm{J}=0$. Like for a single 3JJ qubit, the relevant solutions corresponding to actual minima are readily verified to have
\beq
  \phi_1=\phi_3\;,\qquad\phi_4=\phi_6\;,
\eeql{5to3}
leaving one with three nontrivial extremum equations.

Two solutions are readily found as
\begin{gather}
  \phi_0^\mathrm{FM}=0\;,\qquad\phi_1^\mathrm{FM}=\phi_4^\mathrm{FM}=\pm\arccos\biggl(\frac{1}{2\alpha}\biggr)\;,\label{phiFM}\\
  U_\mathrm{J}(\bm{\phi}^\mathrm{FM})=-E_0-\biggl(\frac{1}{\alpha}+2\alpha\biggr)E\;.
\end{gather}
Clearly, these correspond to ferromagnetic (FM) configurations, in which the sense of supercurrent rotation is the same in both loops. In this symmetric device, the currents in the central leg therefore cancel, so there is no phase difference across this leg and the remaining junctions are in the same state as for degenerately biased free 3JJ qubits.

On the other hand, for the antiferromagnetic (AF) configurations, one has
\beq
  \phi_4^\mathrm{AF}=-\phi_1^\mathrm{AF}\;.
\eeql{4m1}
The remaining equations
\begin{subequations}
\begin{align}
  E_0\sin\phi_0^\mathrm{AF}&=2E\sin\phi_1^\mathrm{AF}\;,\label{phi0}\\
  \sin\phi_1^\mathrm{AF}&=\alpha\sin(2\phi_1^\mathrm{AF}{+}\phi_0^\mathrm{AF})\label{phi1}
\end{align}
\end{subequations}%
have to be solved perturbatively in~$E/E_0$. Since the lowest order corresponds to free qubits, one has
\beq
  \phi_1^\mathrm{AF}=\pm\arccos\biggl(\frac{1}{2\alpha}\biggr)+\phi_1^{(1)}\frac{E}{E_0}+\Oc[{(E/E_0)}^2]\;.
\eeql{phi1AF}
Substitution into Eq.~(\ref{phi0}) readily yields
\beq
  \phi_0^\mathrm{AF}=\pm2\frac{E}{E_0}\sqrt{1-\frac{1}{4\alpha^2}}+\Oc[{(E/E_0)}^2]
           \equiv\pm\frac{2(I_\mathrm{p}/2e)}{E_0}+\Oc[{(E/E_0)}^2]\;,
\eeq
in terms of the classical equilibrium persistent current~$I_\mathrm{p}$ ($\hbar=1$). Subsequently, Eq.~(\ref{phi1}) leads to $\phi_1^{(1)}=\pm[(2{-}4\alpha^2)/(4\alpha^2{-}1)]I_\mathrm{p}/2eE$. However, this cancels in the expansion (around a potential minimum) of Eq.~(\ref{UJ}), and one finds
\beq
  U_\mathrm{J}(\bm{\phi}^\mathrm{AF})=U_\mathrm{J}(\bm{\phi}^\mathrm{FM})-\frac{I_\mathrm{p}^2}{2e^2E_0}+\Oc[{(E/E_0)}^2]\;,
\eeql{AF}
an AF coupling equivalent to a mutual inductance $M_\mathrm{eff}=1/4e^2E_0$---precisely the Josephson inductance of the coupling junction. It should be feasible to manufacture devices with $M_\mathrm{eff}$ ranging from typical magnetic values to values corresponding to a dimensionless coupling of order one.

The analogy to the magnetic case~\cite{H_2qb} is complete: in Eq.~(\ref{AF}), the energy \emph{in}crease (over the FM state and to leading order) $\frac{1}{2}E_0(\phi_0^\mathrm{AF})^2=I_\mathrm{p}^2/2e^2E_0$ in the central junction is overcompensated by the \emph{de}crease in energy of the ``qubit'' junctions $2\phi_0^\mathrm{AF}\pt_{\phi_\mathrm{x}} U_\mathrm{J}^\mathrm{min}$, where $\pt_{\phi_\mathrm{x}} U_\mathrm{J}^\mathrm{min}=\mp I_\mathrm{p}/2e$. In fact, the coupling's AF character can be understood without doing the detailed expansion (and, hence, without requiring $E_0$ to be large). Consider $\bm{\phi}'=(\phi_0^\mathrm{FM},\phi_1^\mathrm{FM},\phi_3^\mathrm{FM},-\phi_4^\mathrm{FM},-\phi_6^\mathrm{FM})$; in this perhaps counter-intuitive AF state, the two loop currents have opposite senses, yet there is no phase drop over the central junction. However, this merely means that the state is non-stationary (the T-shaped islands get charged), but it is a well-defined point in the potential landscape, with $U_\mathrm{J}(\bm{\phi}')=U_\mathrm{J}(\bm{\phi}^\mathrm{FM})$ by Eq.~(\ref{UJ}). Therefore, one necessarily has $U_\mathrm{J}(\bm{\phi}^\mathrm{AF})<U_\mathrm{J}(\bm{\phi}^\mathrm{FM})$ for the AF \emph{minimum} state, as is also seen numerically.

\bigskip

For added flexibility, let us consider current-biasing the central leg~\cite{lantz}. This can be described by the potential
\beq\begin{split}
  U(\bm{\phi})&=-E_0\cos\phi_0-\Ic\phi_0-E[\cos\phi_1+\cos\phi_3
    +\alpha\cos(\phi_\mathrm{x}^a{-}\phi_1{-}\phi_3{-}\phi_0)\\
   &\quad +\cos\phi_4+\cos\phi_6+\alpha\cos(\phi_\mathrm{x}^b{-}\phi_4{-}\phi_6{+}\phi_0)]\;,
\end{split}\eeql{UI}
where the phase frustrations will be specified shortly. The bias current is $I_\mathrm{x}=2e\Ic$; it may be large compared to the loop currents but shouldn't exceed the critical one, so $|\Ic|<E_0$. It may look asymmetric to couple the bias to the middle leg only, but the phases in the different legs are not independent. Pending a dynamic analysis of the full Hamiltonian, here we simply use (\ref{UI}) to see what it predicts. Again minimizing w.r.t.\ the phases, one sees that (\ref{5to3}) still applies. As expected, the central leg has to carry most of the bias, viz.,
\beq
  \phi_0=\arcsin\biggl(\frac{\Ic}{E_0}\biggr)+\phi_0^{(1)}\frac{E}{E_0}
  +\phi_0^{(2)}\frac{E^2}{E_0^2}+\Oc[(E/E_0)^3]\;;
\eeql{phi-bias}
it is most consistent to retain $\phi_0^{(2)}$ when evaluating the (large) first two terms of (\ref{UI}) to $\Oc(E/E_0)$, although its actual value cancels. The need to have a stable solution unfortunately limits us to $|\phi_0|<\pi/2$ in (\ref{phi-bias}).

To leading order, the outer arms should behave like degenerate 3JJ qubits, so we have to compensate for the contribution of $\phi_0$ to the total phase bias:
\beq
  \phi_\mathrm{x}^a=\pi+\arcsin\biggl(\frac{\Ic}{E_0}\biggr)\;,\qquad
  \phi_\mathrm{x}^b=\pi-\arcsin\biggl(\frac{\Ic}{E_0}\biggr)\;.
\eeql{compensate}
In the FM configurations, the loop currents again cancel: ${(\phi_0^{(1)})}^\mathrm{FM}={(\phi_0^{(2)})}^\mathrm{FM}=0$, while $\phi_{1,4}^\mathrm{FM}$ are still given by~(\ref{phiFM}). For the AF ones, $\phi_{1,4}^\mathrm{AF}$ are as in (\ref{4m1}) and (\ref{phi1AF}) (though with a different~$\phi_{1,4}^{(1)}$), yielding
\beq
  {(\phi_0^{(1)})}^\mathrm{AF}=\pm\frac{2(I_\mathrm{p}/2eE)}{\sqrt{1-(\Ic/E_0)^2}}\;.
\eeq
Using these to expand the potential (\ref{UI}), one finds the proper generalization of (\ref{AF}),
\beq
  U_\mathrm{J}(\bm{\phi}^\mathrm{AF})=U_\mathrm{J}(\bm{\phi}^\mathrm{FM})
  -\frac{I_\mathrm{p}^2}{2e^2\sqrt{E_0^2-\Ic^2}}+\Oc[(E/E_0)^2]\;.
\eeql{AF-bias}
Increasing the current bias decreases the effective~$E_0$, \emph{in}creasing the coupling energy: as expected, a near-critically biased central junction is more sensitive to changes in current direction in the outer arms. Thus, tunability of the coupling strength (but only up) as in (\ref{AF-bias}) requires an extra current-bias lead for $\Ic$ itself, plus the generation of a flux-bias asymmetry as in (\ref{compensate}). While the latter is a common resource in two-flux-qubit experiments~\cite{galv-expt,IMT2qb}, presently a substantial relative detuning of the coupling strength would require a flux-bias asymmetry of a significant fraction of a flux quantum.

In practical terms, the prediction is an interaction Hamiltonian $H_\mathrm{int}=J\sigma_z^a\sigma_z^b$, with coupling $J=(\hbar/2e)I_\mathrm{p}^2/\sqrt{I_\mathrm{c0}^2{-}I_\mathrm{b}^2}$, in terms of the critical current of the shared junction $I_\mathrm{c0}$ and the bias $I_\mathrm{b}$. Equivalently, the effective AF mutual inductance has absolute value $M_\mathrm{eff}=(\hbar/2e)/\sqrt{I_\mathrm{c0}^2{-}I_\mathrm{b}^2}$, which again equals the Josephson inductance of the coupling junction at the working point. Linearizing these relations, we can sketch how an $I_\mathrm{b}$-noise translates into a $J$-noise. One finds the width of the coupling energy as
\beq
  \Delta J=\frac{\hbar I_\mathrm{p}^4I_\mathrm{b}^2}
  {4e^2{(I_\mathrm{c0}^2{-}I_\mathrm{b}^2)}^3}\,S_\mathrm{b}(0)\;,
\eeq
where the bias-noise power is $S_\mathrm{b}(\omega)=\int\!dt\,\e^{i\omega t}\langle\delta I_\mathrm{b}(t)\delta I_\mathrm{b}(0)\rangle$. This concludes the treatment of the coupling; further details depend on the circuit into which it is incorporated.

\bigskip

Generalization to asymmetric devices and/or biases presents no trouble. It seems more interesting to further investigate coupling tunability. The standard ploy of a flux-biased compound central junction seems feasible, though it again does not directly lead to \emph{switchable} coupling, which here would require $E_0\To\infty$ not $E_0\To0$; also, flux leakage to the qubit loops is a possible problem.

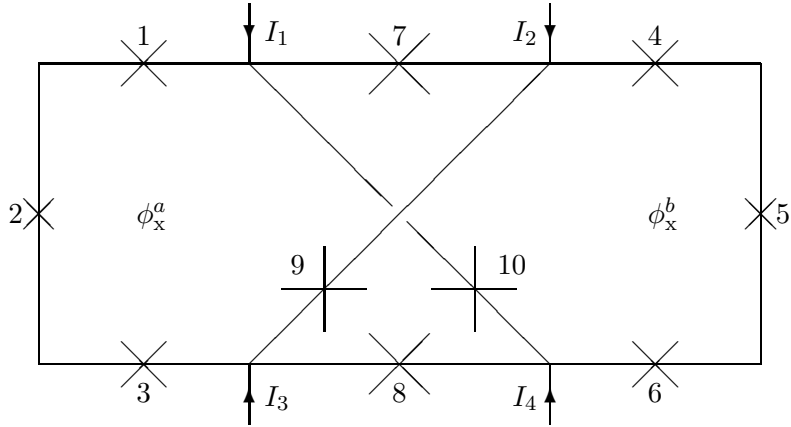
\begin{figure}[ht]
\begin{picture}(110,50)
  \put(7,5){\line(1,0){96}}
  \put(7,45){\line(1,0){96}}
  \put(7,5){\line(0,1){40}}
  \put(103,5){\line(0,1){40}}
  \put(35,5){\line(1,1){40}}
  \put(35,45){\line(1,-1){19}}
  \put(75,5){\line(-1,1){19}}
  \put(51,1){\line(1,1){8}}
  \put(51,9){\line(1,-1){8}}
  \put(51,41){\line(1,1){8}}
  \put(51,49){\line(1,-1){8}}
  \put(39.3,15){\line(1,0){11.3}}
  \put(45,9.3){\line(0,1){11.3}}
  \put(59.3,15){\line(1,0){11.3}}
  \put(65,9.3){\line(0,1){11.3}}
  \put(18,2){\line(1,1){6}}
  \put(18,8){\line(1,-1){6}}
  \put(18,42){\line(1,1){6}}
  \put(18,48){\line(1,-1){6}}
  \put(86,2){\line(1,1){6}}
  \put(86,8){\line(1,-1){6}}
  \put(86,42){\line(1,1){6}}
  \put(86,48){\line(1,-1){6}}
  \put(5,23){\line(1,1){4}}
  \put(5,27){\line(1,-1){4}}
  \put(101,23){\line(1,1){4}}
  \put(101,27){\line(1,-1){4}}
  \put(20,23.8){$\phi_\mathrm{x}^a$}
  \put(88,23.8){$\phi_\mathrm{x}^b$}
  \put(20,47.5){1}
  \put(3,23.8){2}
  \put(20,0){3}
  \put(88,47.5){4}
  \put(105,23.8){5}
  \put(88,0){6}
  \put(54,47.5){7}
  \put(54,0){8}
  \put(40.5,17){9}
  \put(68,17){10}
  \put(35,-3){\line(0,1){8}}
  \put(75,-3){\line(0,1){8}}
  \put(35,53){\line(0,-1){8}}
  \put(75,53){\line(0,-1){8}}
  \thicklines
  \put(35,2){\vector(0,1){0}}
  \put(75,2){\vector(0,1){0}}
  \put(35,48){\vector(0,-1){0}}
  \put(75,48){\vector(0,-1){0}}
  \thinlines
  \put(70,-0.5){$I_4$}
  \put(37,48){$I_1$}
  \put(37,-0.5){$I_3$}
  \put(70,48){$I_2$}
\end{picture}
\caption{A ``crossbar" circuit in which 3JJ qubits are coupled with tunable strength and sign.}
\label{crossbar}
\end{figure}

An attempt in this direction is presented in Fig.~\ref{crossbar}. The ``qubit" junctions 1--6 are taken the same as before, while the ``coupling" junctions 7--10 are all large, with Josephson energy~$E_0$. It is further assumed that no flux is threading the ``crossbar" part of the circuit, that is, the legs which are drawn diagonally for clarity should actually lie almost on top of each other. This fixes $\phi_{10}=\phi_7-\phi_8+\phi_9$ (for a convention in which positive phases correspond to a clockwise current in the $a$-loop). As before, also $\phi_{2,5}$ are readily eliminated, while all stationary configurations have $\phi_1=\phi_3$ and $\phi_4=\phi_6$. In terms of the remaining variables, the potential can be written as
\beq\begin{split}
  U(\bm{\phi})&=-E[2\cos\phi_1+\alpha\cos(\phi_\mathrm{x}^a{-}2\phi_1{-}\phi_7{-}\phi_9)+2\cos\phi_4
  +\alpha\cos(\phi_\mathrm{x}^b{-}2\phi_4{-}\phi_8{+}\phi_9)]\\
  &\quad-E_0[\cos\phi_7+\cos\phi_8+\cos\phi_9+\cos(\phi_7{-}\phi_8{+}\phi_9)]
  -\Ic_1\phi_7-\Ic_4\phi_8+(\Ic_3{+}\Ic_4)\phi_9\;.
\end{split}\eeql{U10}

The counterpart to (\ref{compensate}) becomes that, to leading order in~$E/E_0$, degenerate bias should be implemented as
\beq
  \phi_\mathrm{x}^a-\phi_7^{(0)}-\phi_9^{(0)}=\phi_\mathrm{x}^b-\phi_8^{(0)}+\phi_9^{(0)}=\pi\;.
\eeq

The four bias leads enable a bewildering degree of tunability, subject to $\sum_{j=1}^4\Ic_j=0$. However, only two one-parameter families will be considered, which are thought to be the most useful: (A)~$\Ic_1=-\Ic_2=-\Ic_3=\Ic_4=\Ic$, and (B)~$\Ic_1=\Ic_2=-\Ic_3=-\Ic_4=\Ic$. In biasing scheme (A) one trivially finds $\phi_7^{(0)}=\phi_8^{(0)}=\arcsin(\Ic/E_0)$ and $\phi_9^{(0)}=0$: the external current flows through the horizontal legs. Conversely, in scheme (B), $\phi_7^{(0)}=\phi_8^{(0)}=0$ and $\phi_9^{(0)}=\arcsin(\Ic/E_0)$ so that the bias flows through the vertical legs. Minimizing $U$ in (\ref{U10}), the leading corrections to these phases due to the qubit currents are readily found. Substituting these back into $U$, one obtains the energies of the various stationary configurations.

\emph{Bias scheme (A)} favours the FM states:
\beq
  U^\mathrm{A,AF}-U^\mathrm{A,FM}=
  \frac{I_\mathrm{p}^2}{4e^2}\biggl(\frac{1}{\sqrt{E_0^2-\Ic^2}}-\frac{1}{E_0}\biggr)\;.
\eeql{crossbar-coupl}
The interpretation is clear: the vertical, unbiased junctions mediate an AF interaction as in Fig.~\ref{circuit}. The horizontal junctions mediate a ``twisted" FM interaction, which overcomes the AF one due to the current bias. Compared to (\ref{AF}) and (\ref{AF-bias}), Eq.~(\ref{crossbar-coupl}) is a factor 2 smaller: the two junctions in parallel in Fig.~\ref{crossbar} have a coupling energy which is twice as large, and therefore an effective inductance twice as small, as the single shared junction in Fig.~\ref{circuit}.

On the other hand, \emph{bias scheme (B)} favours the AF states, with a coupling energy which is the exact opposite of the one in~(\ref{crossbar-coupl}). In fact, if one envisions flipping the right part of Fig.~\ref{crossbar} it becomes clear that, when changing from scheme (A) to scheme (B), the role of FM and AF states is simply reversed.

\begin{figure}[ht]
\begin{picture}(110,50)
  \put(7,5){\line(1,0){96}}
  \put(7,45){\line(1,0){96}}
  \put(7,5){\line(0,1){12}}
  \put(7,45){\line(0,-1){12}}
  \put(3,17){\line(0,1){16}}
  \put(11,17){\line(0,1){16}}
  \put(3,17){\line(1,0){8}}
  \put(3,33){\line(1,0){8}}
  \put(103,5){\line(0,1){12}}
  \put(103,45){\line(0,-1){12}}
  \put(99,17){\line(0,1){16}}
  \put(107,17){\line(0,1){16}}
  \put(99,17){\line(1,0){8}}
  \put(99,33){\line(1,0){8}}
  \put(39,5){\line(0,1){40}}
  \put(35,21){\line(1,1){8}}
  \put(35,29){\line(1,-1){8}}
  \put(71,5){\line(0,1){40}}
  \put(67,21){\line(1,1){8}}
  \put(67,29){\line(1,-1){8}}
  \put(52,42){\line(1,1){6}}
  \put(52,48){\line(1,-1){6}}
  \put(22,23.8){$\phi_\mathrm{x}^a$}
  \put(54,23.8){$\phi_\mathrm{x}^b$}
  \put(86,23.8){$\phi_\mathrm{x}^c$}
  \put(5.5,24){$U_a$}
  \put(101.5,24){$U_c$}
  \put(34.5,24){1}
  \put(54,47.5){2}
  \put(74,24){3}
\end{picture}
\caption{Sign-tunable galvanic coupling of ``black-box'' qubits $a$ and $c$ via an intermediate $b$-loop.}
\label{mediated}
\end{figure}
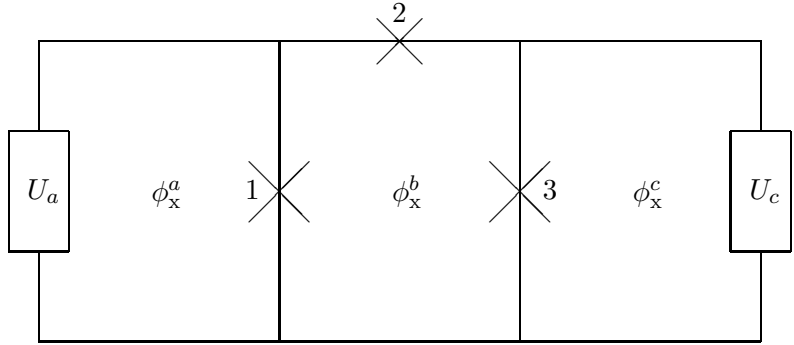

Another scheme for sign-tunable coupling, more similar to the devices described in~\cite{AB,Plourde}, is shown in Fig.~\ref{mediated}. We generalize and at the same time simplify the analysis by taking the two qubits as ``black boxes'', which may be asymmetric, biased away from degeneracy, and/or of a different design than 3JJ. This stresses that the crucial element in the analysis is the coupler, not the qubits. The potential reads
\beq
  U_\mathrm{J}(\bm{\phi})=U_a(\phi_\mathrm{x}^a{-}\phi_1)+U_c(\phi_\mathrm{x}^c{+}\phi_3)
  -E_1\cos\phi_1-E_2\cos(\phi_\mathrm{x}^b{+}\phi_1{-}\phi_3)-E_3\cos\phi_3\;.
\eeql{UJ-med}
Here, the qubit potential $U_a$ is already minimized over its internal degrees of freedom; we denote $U_a'(\phi_\mathrm{x}^a)=-I_{\mathrm{p}a}/2e$ and $U_a''(\phi_\mathrm{x}^a)=\chi_a/4e^2$, where $\chi_a$ is the qubit susceptibility. These have two values depending on the qubit state, with $I_{\mathrm{p}a}$ having two different signs; similarly for the $c$-qubit. The scheme is analogous to its magnetic counterpart in~\cite{A&A}, which means that $E_2$ is a variable of order one. We expand to second order in terms of only $E_1^{-1}$ and $E_3^{-1}$, implementing the coupling \emph{per~se}.

For this purpose, we write $\phi_1=\phi_1^{(1)}+\phi_1^{(2)}$ and $\phi_3=\phi_3^{(1)}+\phi_3^{(2)}$. Minimizing $U_\mathrm{J}$, one finds
\begin{align}
  E_1\phi_1^{(1)}&=-\frac{I_{\mathrm{p}a}}{2e}-E_2\sin\phi_\mathrm{x}^b\;,\\
  E_3\phi_3^{(1)}&=\frac{I_{\mathrm{p}c}}{2e}+E_2\sin\phi_\mathrm{x}^b\;,\\
  E_1\phi_1^{(2)}&=-\phi_1^{(1)}\frac{\chi_a}{4e^2}
    +(\phi_3^{(1)}{-}\phi_1^{(1)})E_2\cos\phi_\mathrm{x}^b\;,\\
  E_3\phi_3^{(2)}&=-\phi_3^{(1)}\frac{\chi_c}{4e^2}
    +(\phi_1^{(1)}{-}\phi_3^{(1)})E_2\cos\phi_\mathrm{x}^b\;.
\end{align}
Substituting these into (\ref{UJ-med}) for $U_\mathrm{J}$ is straightforward. To $\Oc(E_{1,3}^{-2})$ one finds trivial constants, terms $\propto I_{\mathrm{p}a,c}$ representing small shifts in effective bias, terms $\propto I_{\mathrm{p}a,c}^2$ representing small (Josephson) inductances, and some small corrections $\propto\chi_{a,c}$. Most interesting for us is the coupling
\beq
  U_\mathrm{J}=\cdots+\frac{E_2\cos\phi_\mathrm{x}^b}{4e^2E_1E_3}I_{\mathrm{p}a}I_{\mathrm{p}c}
  \equiv H_\mathrm{int}\;.
\eeql{med-res}
This is the expected result: expressing (9) in~\cite{A&A} for the magnetic case as $H_\mathrm{int}=M_{ab}M_{bc}\chi_b^0I_{\mathrm{p}a}I_{\mathrm{p}c}$ (which, incidentally, generalizes to the quantum case), (\ref{med-res}) follows under $M_{ab}\mapsto1/4e^2E_1$, $M_{bc}\mapsto1/4e^2E_3$~\cite{galv-expt}, and $\chi_b^0\mapsto4e^2E_2\cos\phi_\mathrm{x}^b$, the susceptibility of an \mbox{rf-SQUID} of vanishing inductance. Compared to the ``crossbar" design, the proposal of Fig.~\ref{mediated} may be easier to implement. It also avoids realizing a small coupling as the difference of two larger ones, which may be sensitive to control errors. The restriction $E_2\ll E_{1,3}$ in the above can be lifted~\cite{VP}; the corresponding closed-form generalization of (\ref{med-res}) will be presented elsewhere shortly.

The derivations in this note have, hopefully, stressed transparent results and clarity rather than detailed modeling. For the latter, one should first of all study the classical potential nonperturbatively, since real coupling strengths will not be infinitesimal. One can ($i$) calculate the actual (coupling) energies at the codegeneracy point (symmetry point in parameter space) from a potential such as (\ref{UJ}), as is relevant to e.g.\ spectroscopy or at finite temperatures. Alternatively, one can ($ii$) calculate the classical stability diagram $\bm{s}(\bm{\phi}_\mathrm{x})$, indicating which flux state $(s_a,s_b)=(1,-1)$ etc.\ prevails as a function of the biases; this is useful for, e.g., impedance measurement~\cite{galv-expt,IMT2qb}. While ($i$) and ($ii$) are not equivalent outside the range where the energy--bias relation can be linearized, both are special cases of the classical ``band structure'' $U(\bm{s},\bm{\phi}_\mathrm{x})$, which is readily calculated whenever needed.

Also, the classical derivation of (\ref{med-res}) etc.\ is a simplification. A full quantum analysis could be perturbative in the coupling or not, since its relevance depends on the junction capacitances, not on the classical interaction strength. Such an analysis can be transcribed from its magnetic counterpart~\cite{A&A}, but this will not be pursued here.

Extending the above designs to a linear qubit chain looks straightforward; a possible adaptation to 2D qubit lattices remains to be investigated.

\end{document}